\documentclass[10pt]{aa}
\usepackage{graphicx}
\usepackage{natbib}
\bibpunct{(}{)}{;}{a}{}{,}
\begin{document}
\title{Observational detection of eight mutual eclipses and occultations between the satellites of Uranus}
\author {A.~A.~Christou\inst{1} \and F.~Lewis \inst{2,3,4} \and P.~Roche \inst{2,3,4} \and Y.~Hashimoto \inst{5,6} \and D.~O'Donoghue \inst{5} \and H.~Worters \inst{5,6} \and D.~A.~H.~Buckley \inst{5,6} \and T.~Michalowski \inst{7} \and D.~J.~Asher \inst{1} \and A.~Bitsaki \inst{8} \and A.~Psalidas \inst{9} \and V.~Tsamis \inst{10,11} \and K.~N.~Gourgouliatos \inst{12} \and A.~Liakos \inst{13} \and M.~G.~Hidas \inst{3,14} \and T.~M.~Brown \inst{3,14}
}
\institute{
           Armagh Observatory, College Hill,
           Armagh BT61 9DG, Northern Ireland, UK
           e-mail: aac@arm.ac.uk
      \and Faulkes Telescope Project
           School of Physics and Astronomy, Cardiff University,
           Queens Buildings, 5 The Parade, Cardiff CF24 3AA, UK 
      \and Las Cumbres Observatory Global Telescope, 
           6740 Cortona Dr.~Ste.~102, Goleta, CA 93117, USA
      \and Department of Physics and Astronomy, The Open University, 
           Walton Hall, Milton Keynes, MK7 6AA, UK
      \and South African Astronomical Observatory, P.O.~Box 9,
           Observatory, 7935, South Africa
      \and Southern African Large Telescope Foundation, P.O.~Box 9,
           Observatory, 7935, South Africa
      \and Poznan (A.~Mickiewicz Univ.), Astr.~Obs.~of A.~Mickiewicz Univ.,
 	   ul.Sloneczna, PL-60-286 Poznan, Poland
      \and Hellenic-American Educational Foundation, Athens College,
      P.O.~Box 65005, Psychico, 15410 Athens, Greece
      \and Educational Research and Pedagogy Unit, School of Humanities, 
           Hellenic Open University, 23 Sachtouri Str.,
           26222 Patras, Greece
      \and Hellenic Astronomical Association, I.~Metaxa \& Vas.~Pavlou Str., 
           Palaia Penteli Attikis, 15236 Athens, Greece
      \and Ellinogermaniki Agogi School Observatory, Dimitriou Panagea Str.,  
           Pallini Attikis, 15351 Athens, Greece
      \and Institute of Astronomy, University of Cambridge,
           Madingley Road, Cambridge CB3 0HA, UK  
      \and Department of Astrophysics, Astronomy and Mechanics, 
           University of Athens, GR-15784 Zografos,
           Athens, Greece
      \and Department of Physics, University of California, 
           Santa Barbara, CA 93106, USA
          }
\date{Received 16 December 2008 / Accepted 9 February 2009}
\abstract
{}
{We carried out observations, with five different instruments ranging in aperture from 0.4m to 10m, 
of the satellites of Uranus during that planet's 2007 Equinox. Our observations covered specific intervals of time when mutual eclipses 
and occultations were predicted.}
{The observations were carried out in the near-infrared part of the spectrum to mitigate the glare from the planet.
Frames were acquired at rates $>1/\mbox{\rm min}$. Following modelling and subtraction of the planetary source
from these frames, differential aperture photometry was carried out on the satellite pairs involved in the predicted events. 
In all cases but one, nearby bright satellites were used as reference sources.}
{We have obtained fifteen individual lightcurves, eight of which show a clear drop in the flux from the satellite pair, indicating that 
a mutual event took place. Three of these involve the faint satellite Miranda. All eight lightcurves were model-fitted to yield best estimates of 
the time of maximum flux drop and the impact parameter. In three cases best-fit albedo ratios were also derived. We used these estimates 
to generate intersatellite astrometric positions with typical formal uncertainties of $<{0.}^{\arcsec}01$, several times better than conventional 
astrometry of these satellites. The statistics of our estimated event midtimes show a systematic lag, with the observations later than predictions. 
In addition, lightcurves of two partial eclipses of Miranda show no statistically significant evidence of a light drop, at variance with the predictions. 
These indicate that new information about the Uranian satellite system is contained in observations of mutual events acquired here and by other groups.}
{}
\titlerunning{Eclipses and occultations between Uranian satellites}
\authorrunning{Christou, et al.}

\maketitle
\section{Introduction}
The satellites of the planets Jupiter, Saturn and Uranus 
undergo seasons of mutual eclipses and occultations at equinox when the Sun 
and the Earth respectively pass through the planet's equatorial plane. 
These so-called ``mutual events'' yield very precise positional 
measurements of the satellites \citep{Vasundhara.et.al.2003,Noyelles.et.al.2003,EmelyanovGilbert2006}.

The 2007 Uranian equinox presented the first opportunity to observe
mutual events between the classical Uranian satellites: Miranda (V), Ariel (I), Umbriel (II), Titania (III) and Oberon (IV)
\citep{Christou05,Arlot.et.al.2006}. These observations can potentially  
lead to the greatest incremental improvement of the satellite ephemerides 
and system constants since the Voyager 2 flyby of the Uranian system in 1986.

In the present study, we report on the outcome of an observational programme of these mutual events using
five different instruments with 0.4m, 2m and 10m apertures during 2007.
Analysis of one of these observations, which to our knowledge was also the first of a mutual 
event between two Uranian satellites, was reported in \citet{Hidas.et.al.2008} hereafter referred to as HCB08.
In the following section we describe our observational strategy and circumstances for the observed events. 
In Section 3 we explain the methods we employed for data reduction, particularly 
where these differ from the approach used in HCB08. In Section 4 we present our results in the form of lightcurves, 
model-fitted parameter estimates and astrometric offsets for the satellite pairs involved in each event. In addition, we show that a significant 
offset exists between predicted and observed event midtimes. In Section 5 we present our conclusions.

\section{Observations}
Our programme of observations spanned the period 04/05/2007 - 30/11/2007 and employed the instruments 
and facilities listed in Table~\ref{tab:Sites}. The majority of the observations were carried out with the Faulkes Telescopes North (FTN) 
and South (FTS) on the island of Maui, Hawaii and at Siding Spring, Australia respectively. Two events were observed from Athens, Greece: one from 
the Ellinogermaniki Agogi School Observatory (ATH1) and one from the Gerostathopoulion Observatory of the University of Athens (ATH2). Finally, two events were attempted with the Southern African Large Telescope (SALT) in Sutherland, South Africa. Our observations are summarised in Table~\ref{tab:Obs}.

\begin{table*}
\begin{minipage}[t]{17cm}
\renewcommand{\footnoterule}{}  
\begin{center}
\caption{Instruments and detectors used in acquiring the observations reported in this paper.}
\begin{tabular}{cccccc}
\hline\noalign{\smallskip}
Site Abbr&      FTN  (F65)      &         FTS   (E10)        &         ATH1            &         ATH2               &            SALT (A60)    \\
\noalign{\smallskip}
\hline   
\noalign{\smallskip} 
Location &    Haleakala, Maui   &   Siding Spring, Australia &     Athens, Greece      &      Athens, Greece        &  Sutherland, South Africa  \\
Lat (N)  &    20d 42m 27s       &      -31d 16m 22s          &     37d 59m 52s         &       37d 58m 07s          &        -32d 22m 46s        \\
Lon (E)  &   203d 44m 38s       &      149d 04m 14s          &     23d 53m 36s         &       23d 47m 00s          &         20d 48m 30s        \\
Aper     &    200cm             &        200cm               &       40cm              &         40cm               &        1000cm              \\
CCD      &   2k$\times$2k EEV   &  2k$\times$2k EEV          & 1392$\times$1040 ATIK   & 1530$\times$1020 SBIG      &    2 2k$\times$4k  EEV     \\ 
         &      CCD42-40        &     CCD42-40               &        16HR             &        ST-8 XMEI           &    CCD44-82 frame xfer     \\ 
Binning  &    2$\times$2        &       2$\times$2           &        NO               &       2$\times$2           &      2$\times$2            \\
Img scale $\arcsec$/pxl\footnote{ Image scale is after binning, where applicable.}  &    0.28              &            0.28            &        0.31             &         1.03               &        0.28                \\\hline
\end{tabular}
\label{tab:Sites}
\end{center}
\end{minipage}
\end{table*}

\begin{table*}
\begin{minipage}[t]{17cm}
\renewcommand{\footnoterule}{}  
\begin{center}
  \caption{Summary of observational circumstances and results for the mutual event observations attempted in this work.}
\scalebox{1.0}{
 \begin{tabular}{cccccccccccc}\hline\noalign{\smallskip}
Date   &Event & Obs. &  Exp.~Time &                          & Number of     & UT of       & UT of       &   Ref.  & Obs. & Seeing & \\  
(DDMMYY)& Type\footnote{We use the event type notation of \citet{Arlot.et.al.2006}.}& Site\footnote{Observatory abbreviations are given in
Table~\ref{tab:Sites}.} &    (sec)   &           Filter         & Frames & First  Exp. & Last Exp.   &   Sat.  & Result\footnote{Each observation results in either POS(itive) or NEG(ative) detection of a lightcurve.} & ($\arcsec$) & Airmass \\\noalign{\smallskip}\hline \noalign{\smallskip}                                                                                                                      
040507 & 4O2P &  FTS &      3     & sdss $\mbox{i}^{\prime}$ &    150        & 19:02:01.3  & 19:32:02.1  & Titania &  POS &   2.25 & 1.62 \\ 
260707 & 1E5P &  FTS &      2     & sdss $\mbox{i}^{\prime}$ &     43        & 18:56:21.6  & 19:22:21.8  & Titania &  NEG &   1.18 & 1.28 \\  
050807 & 4O2P &  FTN &      2     & sdss $\mbox{i}^{\prime}$ &     53        & 13:32:12.2  & 14:12:08.0  & Titania &  POS &   1.39 & 1.15 \\
060807 & 1O5P &  FTN &      2     & sdss $\mbox{i}^{\prime}$ &     14        & 10:32:57.4  & 10:41:53.4  & Titania &  NEG &   0.82 & 1.32 \\
140807 & 2O4P &  ATH1&     30     & IR72                     &     34        & 01:09:39    & 01:46:14    & Titania &  POS &   4.35 & 1.46 \\
200807 & 5O2P &  FTN &      2     & sdss $\mbox{i}^{\prime}$ &     51        & 13:34:07.0  & 14:01:11.7  &  Ariel  &  NEG &   3.68 & 1.30 \\
220807 & 2E5T &  FTN &      2     & sdss $\mbox{i}^{\prime}$ &    100        & 14:35:06.4  & 15:15:24.0  & Titania &  POS &   1.55 & 1.74 \\
240807 & 1O2P &  FTN &      2     & sdss $\mbox{i}^{\prime}$ &    150        & 12:02:13.8  & 13:00:10.1  &  Oberon &  POS &   1.11 & 1.16 \\
220907 & 1E5P &  ATH2&     10     &  Bessell I               &    260        & 18:01:26    & 18:57:35    & Titania &  NEG &   4.26 & 2.08 \\
051007 & 1O5P &  FTS &      2     & sdss $\mbox{i}^{\prime}$ &     36        & 08:55:33.3  & 09:29:28.7  & Oberon  &  NEG &   1.77 & 1.55 \\
121007 & 4E5T &  FTN &      2     & sdss $\mbox{i}^{\prime}$ &     75        & 09:32:47.3  & 09:59:52.7  & Titania &  POS &   2.35 & 1.23 \\
211007 & 1E2P &  FTS &      2     & sdss $\mbox{i}^{\prime}$ &    100        & 13:53:06.3  & 14:30:07.2  & Titania &  NEG &   1.52 & 1.51 \\
221007 & 1E2C &  FTS &      2     & sdss $\mbox{i}^{\prime}$ &    100        & 13:22:23.7  & 14:00:09.8  & Titania &  NEG &   1.34 & 1.38 \\
291107 & 2E4P & SALT & 0.7, 1, 2  & Bessell I                &    --         & 19:00:44    & 19:42:05    &  --     &  NEG &   --   & 1.26 \\
301107 & 3E4P & SALT &    0.7     & Bessell I                &    4299       & 18:34:47    & 19:28:27    &  Star   &  POS &   --   & 1.22 \\
301107 & 1E5T &  FTN &      2     & sdss $\mbox{i}^{\prime}$ &     38        & 08:32:56.0  & 08:57:37.9  & Titania &  POS &   1.58 & 2.18 \\\hline
\end{tabular}}
 \label{tab:Obs}
\end{center}
\end{minipage}
\end{table*}

To facilitate the observational campaign, predictions were tabulated in advance
by the models of \cite{Christou05} and \cite{Arlot.et.al.2006}. Christou based his predictions on the Voyager-era
GUST86 analytical ephemeris \citep{LaskarJacobson87} which is available in binary format through NASA's Navigation and 
Ancillary Information Facility (NAIF) ftp site (ftp://naif.jpl.nasa.gov/pub/naif/) as SPICE kernel URA027. 
Arlot et al reproduced the results by Christou based on this ephemeris and, in addition, generated a different set of 
predictions using the recently developed numerical ephemeris LA06 \citep{Lainey2008}. 

To ensure that each mutual event would be captured in its entirety (if it occurred) and to satisfy the
requirement, imposed by scheduling restrictions, that the total duration of each observation is kept to a minimum,
we adopted the following strategy: for each event, the published start and end times according to the predictions
by \cite{Christou05} and \cite{Arlot.et.al.2006} were compared. The start (end) times of our observations were then 
determined by choosing the earlier (later) of the two start (end) times and subtracting (adding) a time interval 
equal to three times the difference between the two sets of predictions.  

During the course of these observations, \cite{RushJacobson07} (hereafter RJ07) published a new numerical ephemeris for these satellites, available
through NAIF as SPICE kernel URA083. By incorporating this new kernel into the prediction model by Christou, we generated a new set of 
predictions specifically for the events discussed in this paper. This was done partly to refine our choice of observing interval
for those events that were yet to be observed but also to enable comparisons between GUST86, LA06 and the new ephemeris.    

We also strove to maximise the contrast between the faint satellites and the bright planet while, 
at the same time, sampling the planet at a good signal-to-noise. The latter would allow 
{\it a posteriori} modelling and subtraction from each frame in order to facilitate satellite photometry. 
For FTN and FTS, this meant that our implementation was very similar to that used in HCB08, 
namely short exposures using a Sloan Digital Sky Survey (SDSS) $i^{\prime}$ filter at a relatively 
high cadence ($<$ 1 min). We used 2-sec exposures as opposed to the 3-sec used in HCB08 since the 
new observations were generally acquired at lower airmass. The cadence was lower in the new observations 
(ranging between 30 and 40 sec as opposed to $\sim$14 sec in HCB08) due to the non-availability of a certain mode in the instrument control software. 
The two observations carried out from Athens used a similar strategy, utilising longer exposures in order to reach 
a satisfactory level of signal with these smaller apertures. During the observations on 22/09/07, we attempted 
2$\times$2 binning in order to improve the cadence to $\sim10$ sec and acquire more measurements during the critical period. 
However, this resulted in undersampling the planet and complicated the process of extracting it from the images. Combined 
with the relatively high airmass, it introduced a high level of noise in the reduced data.

The SALT observations were carried out in Slot Mode, where a mask is advanced over the entire chips except for a 
slot just above the frame transfer boundary. Instead of half frame transfers at the end of each exposure, 
144 rows are moved (the ``slot''). This allowed exposure times as short as 0.7 seconds. In addition, the
slot field of view was aligned with the satellite orbital plane projected on the sky. The image of Uranus was placed in the 1.5mm 
(equivalent to 100 pixels) gap between the two chips in order to minimise the effects of planetary glare on the satellites.  
Observations on the first night (29/11/07) were carried out through cloud and, although the satellites were detectable on the images,
the signal-to-noise ratio was too low and the sky conditions too variable to allow useful photometry of the event.
On the second night, the telescope enclosure was buffeted by wind and this resulted in the light from the satellites
spreading over a relatively large area of the CCD during the exposures. A nearby star was used as a reference source as all 
the other satellites were on the other chip. A mutual eclipse was successfully detected in this case.

\section{Data Reduction}
The images were reduced using the same method as in HCB08, namely subtraction of the planetary source
followed by differential photometry between the satellite pair involved in the occultation or eclipse
relative to a bright reference satellite, usually Titania. At the end of this process, the lightcurves were 
visually inspected to reveal whether a mutual event, as indicated by a dip, actually took place.
Eight of our lightcurves showed convincing evidence of such a dip and were chosen for further analysis. 
These data are available from the authors upon request.
It is worth noting at this point that, since we cannot separate the signal from the two satellites involved 
in an eclipse, we need to model their combined brightness. Hence, unlike the case in \cite{Arlot.et.al.2008},
our eclipse modelling, to be discussed shortly, depends on their albedo ratio.

\begin{figure}
\resizebox{\hsize}{!}{\includegraphics{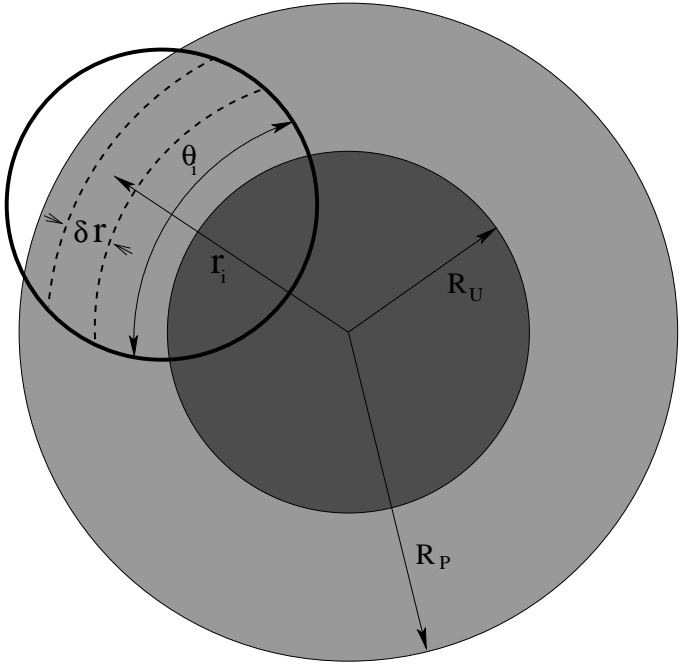}}
 \caption[ ]{Representation, on the impact plane (see text), of the model used to fit our mutual satellite eclipse data. 
             The light lost from the eclipsed satellite (bold circle) is composed of the invisible part of the 
             satellite in the umbra (circle of radius ${R_{\rm U}}$) and the fraction of light lost from each sector $i$
             of thickness $\delta r$ and angular width $\theta_{i}$ in the penumbra (ring of radius $R_{{\rm P}} - R_{{\rm U}}$) 
             at a distance $r_{i}$ from the centre of the umbral cone. For clarity, the penumbra is depicted here as uniformly bright 
             whereas in the model it is of radially-increasing brightness. The index $i$ spans the area fraction of the satellite in the penumbra.}
\label{fig:implane}
\end{figure}

The next stage in the analysis requires a model to estimate the time $t_{\rm min}$ and distance $b$ of closest approach between the two satellites 
on the impact plane as well as the ratio $A$ of their albedos. We define the impact plane as the plane that (a) is perpendicular to the line 
connecting the Sun/Earth with the eclipsing/occulting satellite and (b) contains the uranicentric position of the eclipsed/occulted satellite 
at maximum eclipse/occultation respectively. Our occultation model is the same as the one used in HCB08; here we adopt the new terminology 
$t_{\rm min}$, $b$ and $A$ instead of $t_{o}$, $x$ and $a$ used in that paper. For the eclipses the same terminology is used, with the difference 
that now the eclipsing satellite is replaced by the intersection of its shadow cone with the impact plane. For the purposes of this work, the 
part of the eclipsed satellite in the penumbra is divided into steps of thickness $\Delta r$ and angular size 
$\theta_{\rm i}$ residing at a distance $r_{\rm i}$ from the centre of the shadow cone (see Fig.~\ref{fig:implane}).

Solar illumination within each one of these steps is considered to be constant and equal to $L_{\rm i}$, the fraction of the 
``virtual'' or ``reduced'' sun \citep{Aksnes74,AksnesFranklin76} that is not obstructed by the eclipsing satellite as seen from the eclipsed satellite. 
The ``darkness'' contributions $1 - L_{\rm i}$ from each step are then multiplied by the step areas $ r_{\rm i}\delta r  {\theta}_{\rm i}$ and the (assumed uniform) satellite albedo to estimate the penumbral 
contribution to the loss of light. It is then added to the umbral contribution (equal to the satellite area within the umbral cone) to estimate 
the total dimming of the eclipsed satellite at a given instant in time. 

For eclipses as well as for occultations, we do not take into account surface scattering (``limb darkening'') or phase effects. The latter is less 
than ${3}^{\circ}$ in all cases and affects principally the determination of the impact parameter by a few tens of km or less \citep{Arlot.et.al.2008}.
Similarly, we do not take into account the light travel time between the two satellites which affects the determination of $t_{\rm min}$ by a few seconds
(see eg Noyelles et al.~2003). 

\section{Results} 
\subsection{Positive observations}
\begin{figure*}
\centering
\includegraphics[width=18cm]{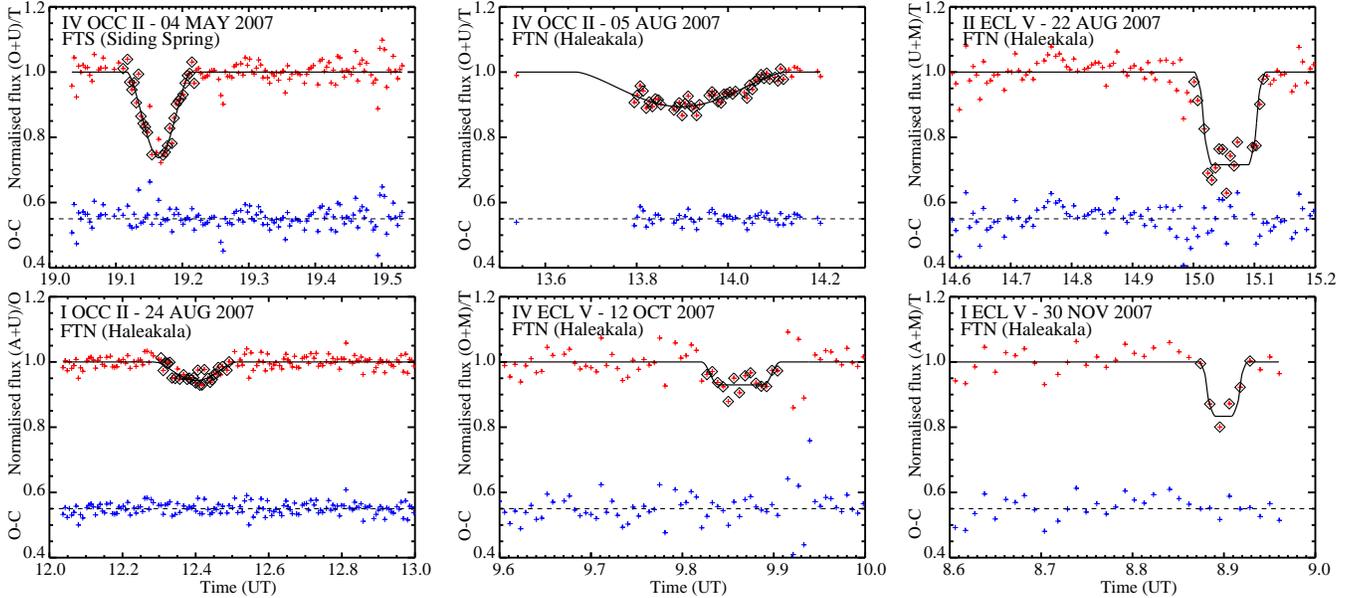}
\caption[ ]{Model fit (black curve) to observations (red ``+'' signs) acquired from Faulkes Telescopes North (FTN) and South (FTS). Diamonds indicate
the data points used in the fit. Residuals (blue ``+'' signs) are referenced to the datum provided by the dashed horizontal line.}
\label{fig:faulkes}
\end{figure*}

\begin{figure*}
\centering
\includegraphics[width=8.5cm]{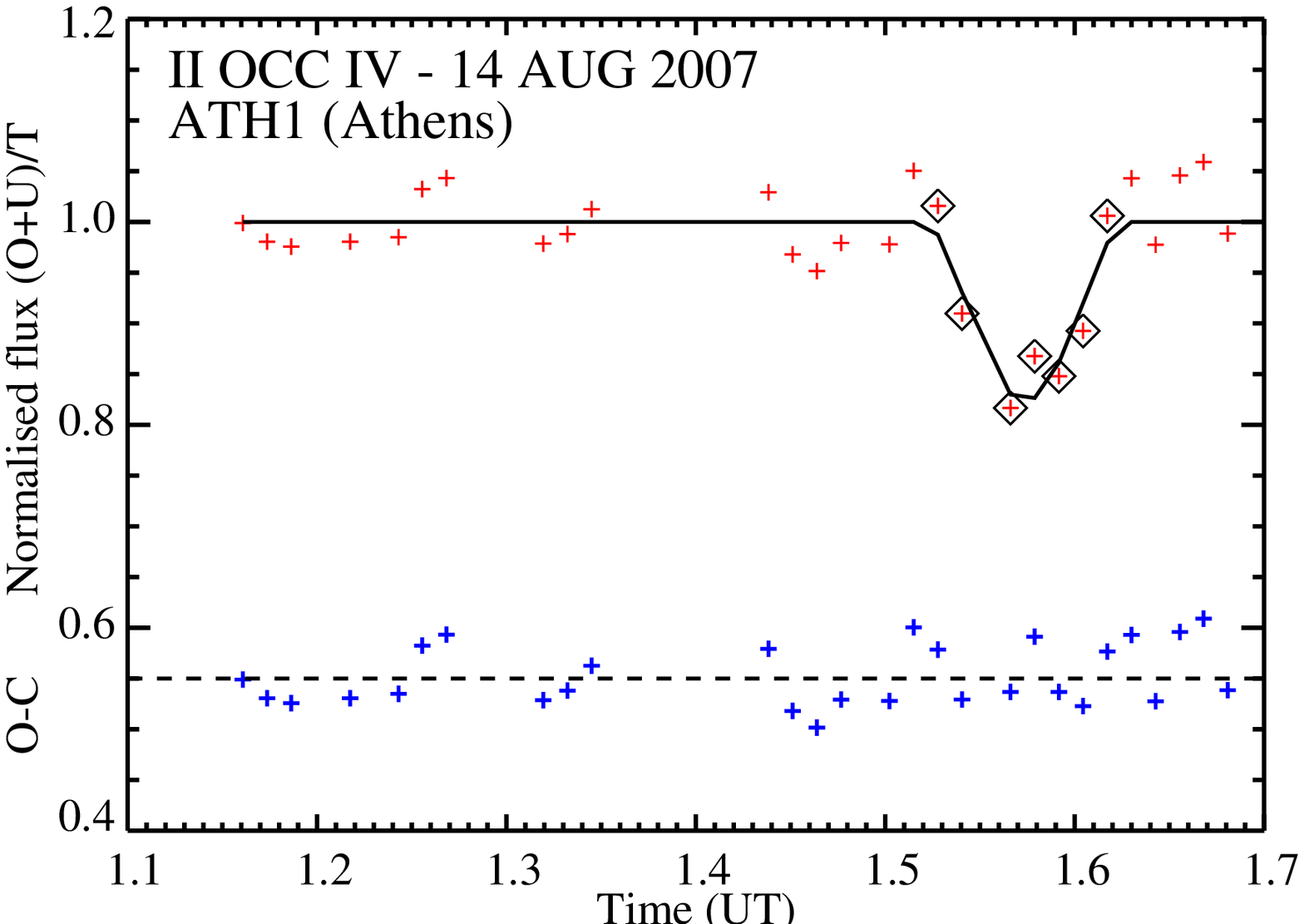}
\includegraphics[width=8.5cm]{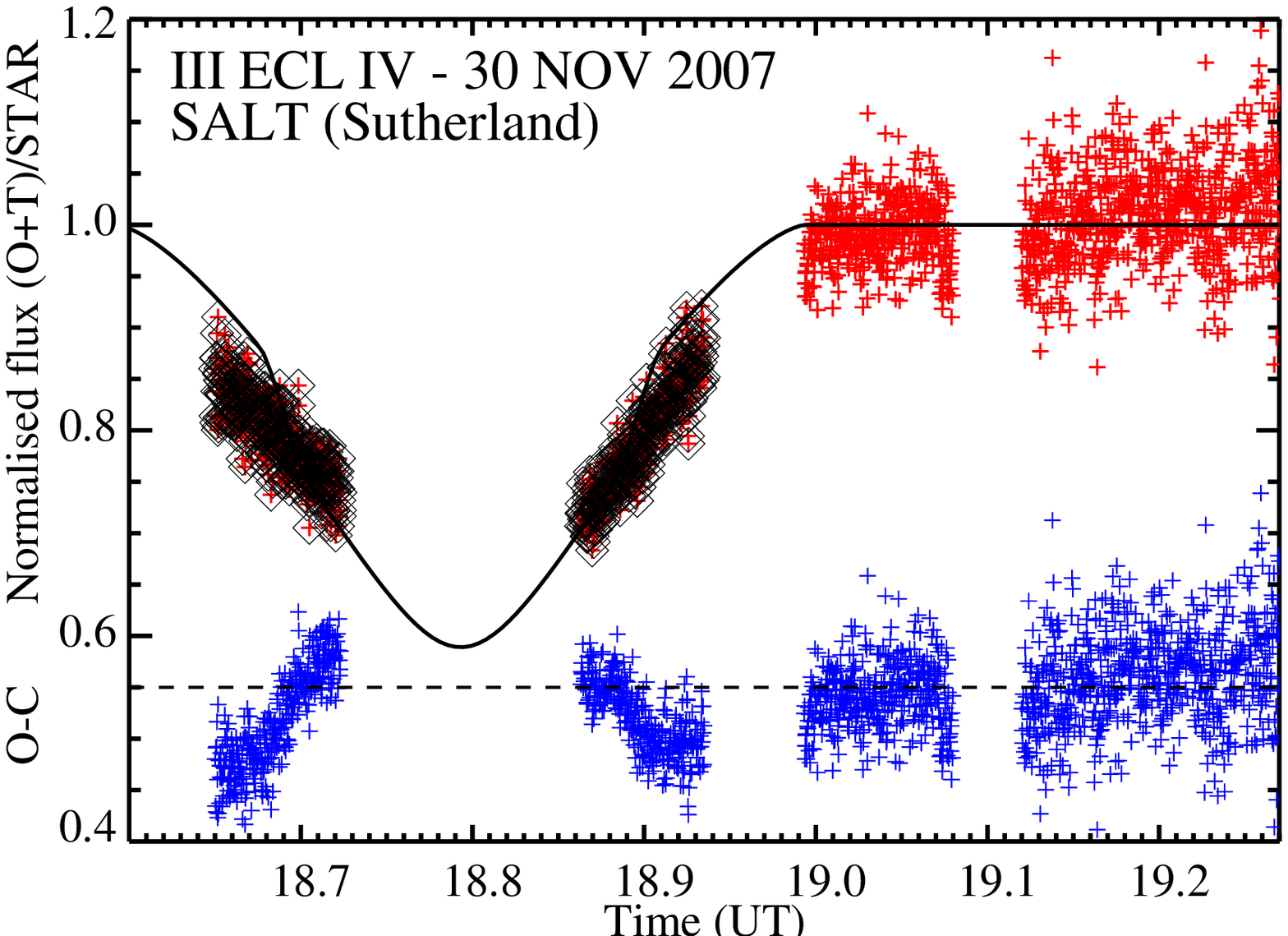}
 \caption[ ]{Model fit to observations acquired from Athens (ATH1; left panel) and Sutherland (SALT; right panel). Notation is as in Fig.~\ref{fig:faulkes}.}
\label{fig:saltathens}
\end{figure*}

During model fitting we considered the relative satellite velocity on the sky plane to be fixed and equal to the one given by RJ07 (available
through the HORIZONS ephemeris service; \citet{Giorgini.et.al.1996}) .
Initial estimates for the ratio $A$ of the satellite albedos were calculated from Table V of \cite{Karkoschka01} by linear interpolation
among the values nearest to the phase angles and wavelengths applicable to each event. This wavelength of ``peak sensitivity'' was 
assumed to be 770 nm (the same of that of the sdss ${i}^{\prime}$ filter used by FTN and FTS) in all cases. Due to the high degree 
of correlation between the albedo ratio and impact parameter in fitting occultations (see HCB08), the albedo ratio $A$ was kept fixed at 
its initial value and only $t_{{\rm min}}$ and $b$ were allowed to vary. On the other hand, simultaneous fitting of 
$t_{{\rm min}}$, $b$ and $A$ was possible for three out of the four eclipse lightcurves. To maintain consistency, we also 
re-reduced the 4th May event reported in HCB08 with an updated set of {\it a priori} parameters. 
The fit results for all eight events are shown in Table~\ref{tab:best_fit}. Numbers shown in brackets are albedo ratio estimates from \cite{Karkoschka01} used 
here as starting values for the fit. All estimated errors are formal 1-$\sigma$ uncertainties. Associated lightcurves and best-fit models are shown separately 
for the FTN/FTS (Fig.~\ref{fig:faulkes}) and  SALT/Athens (Fig.~\ref{fig:saltathens}) observations. Typical root-mean-square (RMS) of the modelled vs observed 
data points ranged from 2\% to 4\% with the exception of the 2O4P event on 14/08/2007 for which a fit RMS of 8\% was obtained.
The SALT data of 30/11/2007 show evidence of significant systematic errors, likely due to marginal observing conditions and the lack of a bright 
reference source on the chip that contained the Oberon/Titania pair. Hence we have kept $A$ fixed and only allowed particular segments of the 
data to be considered in the fitting process in order to enable an acceptable fit. 
 
 \begin{table*}
 \begin{center}
\caption{Best-fit estimates of the parameters of the mutual events successfully observed in this work.} 
 \begin{tabular}{cccccccc}\hline\noalign{\smallskip}
 Date  & Event &  Obs. &    Midtime                 & Impact Parameter     & Albedo                            &  Relative Speed on   & Mean RMS \\
 (DDMMYY) &  Type &  Site &      (UT)                  &        (km)          & Ratio                             &  impact plane (km/s)  & of fit \\
 \noalign{\smallskip}
 \hline                                                                                                                               
 \noalign{\smallskip}                
 040507 & 4O2P &  FTS  & 19:09:53 $\pm$ 3   & 400 $\pm$ 70       &  $1.229 {}^{\mbox{\rm a}} \pm 0.1$      &  $7.078{}^{\mbox{\rm a}}$ & 0.031 \\ 
 050807 & 4O2P &  FTN  & 13:53:49 $\pm$ 30    & 850 $\pm$ 30        &  $1.238{}^{\mbox{\rm a}} \pm 0.1$      &  $1.225{}^{\mbox{\rm a}}$ & 0.019 \\
 140807 & 2O4P &  ATH1 & 01:34:25 $\pm$ 28    & 750 $\pm$ 160      &  $0.813{}^{\mbox{\rm a}} \pm 0.1$      &  $5.765{}^{\mbox{\rm a}}$ & 0.083 \\ 
 220807 & 2E5T &  FTN  & 15:03:37 $\pm$ 5     & 0 $\pm$ 60           &  $0.410 \pm 0.025$  (0.617)       &  $3.647{}^{\mbox{\rm a}}$ & 0.039 \\ 
 240807 & 1O2P &  FTN  & 12:24:04 $\pm$ 30    & 840 $\pm$ 30       &  $1.833{}^{\mbox{\rm a}} \pm 0.1$      &  $2.210{}^{\mbox{\rm a}}$ & 0.018 \\    
 121007 & 4E5T &  FTN  & 09:51:53 $\pm$ 2   & 410${}^{+150}_{-410}$  &  $1.271^{+0.55}_{-0.28}$ (0.759)  &  $5.179{}^{\mbox{\rm a}}$ & 0.042 \\
 301107 & 1E5T &  FTN  & 08:53:58 $\pm$ 2   & 310${}^{+90}_{-130}$ &  $0.830^{+0.22}_{-0.14}$ (1.216)    &  $7.403{}^{\mbox{\rm a}}$ & 0.040 \\
 301107 & 3E4P &  SALT & 18:47:36 $\pm$ 5   & 260 $\pm$ 150    &  $1.113{}^{\mbox{\rm a}} \pm 0.1$      &  $1.992{}^{\mbox{\rm a}}$ & 0.032 \\
 \noalign{\smallskip}
 \hline
 \noalign{\smallskip}
 \end{tabular}
\flushleft
${}^{\mbox{\rm a}}$ These parameters have been assumed and kept fixed during the fitting process. 
 \normalsize
\label{tab:best_fit}
\end{center}
\end{table*}

\begin{table*}
\begin{center}
\caption{Calculated vs observed (C-O) midtime residuals for the ephemerides considered in this paper.}
\begin{tabular}{ccccccc}\hline\noalign{\smallskip}
 Date  &Event & Obs. & Fitted midtime &  C-O sec (km)        &  C-O sec (km)       &   C-O sec (km)\\
(DDMMYY)  & Type & Site &      (UT)      & {(GUST86)}     & {(LA06)}     &  {(RJ07)} \\
\noalign{\smallskip}
\hline 
\noalign{\smallskip}                
040507 & 4O2P &  FTS & 19:09:53  & $-$142 ($-$1010) &  $-$16  ($-$110)  & $-$34 ($-$240) \\
050807 & 4O2P &  FTN & 13:53:49  & $-$149 ($-$180)  &  +137 (+170)  & +19 (+20) \\
140807 & 2O4P & ATH1 & 01:34:25  & $-$177 ($-$1020) &  $-$40  ($-$230)  & $-$77 ($-$440) \\
220807 & 2E5T &  FTN & 15:03:37  & $-$148 ($-$540)  &  $-$47  ($-$170)  & $-$33 ($-$120) \\
240807 & 1O2P &  FTN & 12:24:04  & $-$197 ($-$440)  &  $-$81  ($-$180)  & $-$57 ($-$130) \\
121007 & 4E5T &  FTN & 09:51:53  & $-$117 ($-$610)  &  $-$42  ($-$220)  & $-$77 ($-$400) \\
301107 & 1E5T &  FTN & 08:53:58  & $-$24  ($-$180)  &  $-$35  ($-$260)  & $-$31 ($-$230) \\
301107 & 3E4P & SALT & 18:47:36  & $-$124 ($-$250)  &  +69  (+140)  & $-$34 ($-$70)  \\
\noalign{\smallskip}
\hline
\noalign{\smallskip}
\multicolumn{4}{c}{{Median C-O}}       &  {$-$145} ($-$480) &  {$-$38} ($-$180)   & {$-$34} ($-$180)  \\
\noalign{\smallskip}
\hline
\noalign{\smallskip}
\end{tabular}
\label{tab:midtimes}
\end{center}
\end{table*}

Some interesting features appear in this set of estimates, particularly those for the eclipses. A central (i.e.~$b=0$) configuration for the 4E5T event 
observed on 12/10/2007 cannot be rejected at the $1-\sigma$ level. The initial estimate of $A$ for that event is just outside the
$3-\sigma$ contour around the best-fit value in the $A$ vs $b$ plot of the $\chi^{2}$ goodness-of-fit statistic (see HCB08 for details). 
It is well outside the $3 - \sigma$ contour for the 2E5T event observed on the 22/08/2007 and lies on the $2-\sigma$ contour for the 1E5T event observed
on the 30/11/2007. One explanation for this discrepancy between predictions and observations could be that, since the
satellite hemispheres visible to the Earth at the time of the observations contained areas not in view during the 1997 observations used by 
\cite{Karkoschka01}, we are seeing the effects of albedo variations that have only recently been illuminated by the Sun. Alternatively,
our formal uncertainties may have underestimated the actual photometric errors, for example if the proximity of Miranda to Uranus 
somehow introduced systematics in our measurements.  

These data can be used to gauge the precision of the available ephemerides of the satellites.
Although definitive statements on this issue must await future analysis of the complete set of observations acquired by various groups 
in the course of this international campaign, we can take advantage of the relatively large number of events 
reported in this work by looking at the statistics of the best-determined parameter, namely the midtime $t_{{\rm min}}$. 
In Table~\ref{tab:midtimes} we present the offsets between the observed midtimes and those predicted by models using GUST86, LA06
and RJ07. The numbers in brackets provide the same quantities in km using the speeds from Table~\ref{tab:best_fit}. 
It is evident that the two latter ephemerides performed a factor of $3-4$ better than the Voyager-era GUST86 in
reproducing the satellite-to-satellite positions (along their relative sky velocity vector) to 180 km or 0.013 arcsec.
It is also apparent that the LA06 and RJ07 offsets are not randomly distributed between positive and negative values but 
instead indicate that the observations lag behind the predictions by a few hundred km. This finding is in accordance
with the observations of a single eclipse of Titania by Umbriel reported by \cite{Arlot.et.al.2008} and four events involving 
Ariel, Umbriel, Titania and Oberon reported by \cite{MillerChanover08}. 
We view this as evidence that new information about the Uranian satellite system is contained in the acquired measurements, in accordance with 
the expectations of \citet{Christou05} and \citet{Arlot.et.al.2006}.
  
The final step in our analysis is the conversion of our fitted set of parameters to satellite positions in a celestial
coordinate system. For these purposes the necessary frame rotations were calculated using vectors from RJ07 available through HORIZONS. 
In Table~\ref{tab:residuals} we provide the results in the form of (a) intersatellite positions - occulted (eclipsed) satellite relative to the occulting (eclipsing) satellite - in the J2000 Earth
Equatorial frame at time $t_{\rm min}$ (Columns 4 and 5); and (b) offsets between these positions and those generated by RJ07 at the same time
as $t_{\rm min}$ (Columns 6 and 7). 
It is important to note that these are impact plane coordinates; in other words, while the positions derived for
occultations are Earth-centred, those derived from eclipses are Sun-centred with $t_{\rm min}$ referring to the moment of maximum eclipse 
as observed from the {\it Sun}. Hence these times are different than those given in Tables~\ref{tab:best_fit} and \ref{tab:midtimes}.     
  
\begin{table*}
\begin{center}
\caption{Satellite-satellite relative positions and O-C residuals on the impact plane as determined from our observations in equatorial J2000 coordinates.}
 \begin{tabular}{cccccccc}\hline\noalign{\smallskip}
 Date  &Event & Fitted midtime & \multicolumn{2}{c}{Relative Position ($\arcsec$)}        &  \multicolumn{2}{c}{O-C ($\arcsec$)}\\
 (DDMMYY)  & Type &      (UT)      & $\Delta \alpha \cos \delta $  &            $\Delta \delta$    &  $\Delta \alpha \cos \delta$ & $\Delta \delta$ \\
\noalign{\smallskip}
\hline 
\noalign{\smallskip}                
040507 & 4O2P &  $19$:$09$:$53$ & $-$0.0256 ($\pm$0.0037) &  $-$0.0069 ($\pm$0.0015) & +0.0042 & +0.0181 \\
050807 & 4O2P &  $13$:$53$:$49$ & $-$0.0587 ($\pm$0.0024) &  $-$0.0167 ($\pm$0.0026) & +0.0070 & +0.0003 \\
140807 & 2O4P &  $01$:$34$:$25$ & $-$0.0517 ($\pm$0.0112) &  $-$0.0141 ($\pm$0.0116) & +0.0070 & $-$0.0319 \\
220807 & 2E5T &  $15$:$11$:$37^{\ast}$ & 0.0000 ($\pm$0.0031) & 0.0000 ($\pm$0.0013) & $-$0.0149 & +0.0030 \\
240807 & 1O2P &  $12$:$24$:$04$ & +0.0585 ($\pm$0.0019) &  +0.0157 ($\pm$0.0041) & $-$0.0012 & +0.0097 \\
121007 & 4E5T &  $09$:$58$:$53^{\ast}$ & +0.0280 (${}^{+0.0052}_{-0.0059}$) & $-$0.0025 ($\pm$0.0005) & +0.0012 & +0.0205 \\
301107 & 1E5T &  $08$:$55$:$10^{\ast}$ & $-$0.0208 (${}^{+0.0036}_{-0.0033}$) & +0.0044 ($\pm$0.0008) & $-$0.0044 & +0.0244 \\
301107 & 3E4P &  $18$:$48$:$44^{\ast}$ &  +0.0178 ($\pm$0.0046) & $-$0.0026 ($\pm$0.0007) & $-$0.0106 & $-$0.0046 \\
\noalign{\smallskip}
\hline
\noalign{\smallskip}
\end{tabular}

${}^{\ast}$ Midtime of the event as observed from the centre of the Sun. Earth midtimes for these events are given in Table \ref{tab:best_fit}.
 \label{tab:residuals}
\end{center}
\end{table*}

\subsection{Negative Observations}
\begin{table*}
\begin{center}
  \caption{Circumstances of those observations which gave negative results.}
 \begin{tabular}{cccccccc}\hline\noalign{\smallskip}
 Date   &Event & Obs. &     & Meas. &  \multicolumn{3}{c}{Predicted Flux Drop}\\
 (DDMMYY) & Type & Site & Cov. &  rms & {(GUST86)}     & {(LA06)}     &  {(RJ07)} \\
\noalign{\smallskip}
\hline 
\noalign{\smallskip}                
260707 & 1E5P &  FTS & 1.0  & 0.021 & 0.111  & 0.048  & 0.050 \\
060807 & 1O5P &  FTN & 1.0  & 0.022 & 0.126  & 0.130  & 0.154 \\
200807 & 5O2P &  FTN & 0.7  & 0.130 & 0.050  & 0.103  & 0.087 \\
220907 & 1E5P & ATH2 & 1.0  & 0.224 & 0.130  & 0.080  & 0.086 \\
211007 & 1E2P &  FTS & 1.0  & 0.030 & 0.033  & ---    & 0.007 \\
221007 & 1E2C &  FTS & ---  & 0.043 & ---    & ---    & ---   \\
\noalign{\smallskip}
\hline
\noalign{\smallskip}
\end{tabular}

Dashes (``---'') indicate that the event was expected to be a miss by the respective model.
 \label{tab:negatives}
\end{center}
\end{table*}

Seven of our observations resulted in lightcurves that did not show obvious signs of a photometric dip. In the case of the 051007 observation,
image acquisition began too late to cover the main interval of interest fully or partially. For the remaining six, a dip was 
expected to occur at various degrees of confidence. These are summarised in Table~\ref{tab:negatives}. Column 4 gives the 
fraction of the event, as predicted by RJ07, that overlapped with the time interval covered by our observations. 
Column 5 gives the statistical variation of our photometric measurements about the mean, normalised to 1, for each observation. 
The three last columns give the expected R-band flux drop according to GUST86, LA06 and RJ07 respectively. 

We can gauge the significance of those negatives by using the criterion of \cite{Birlan.et.al.2008}. 
Those authors concluded that a brightness drop of a magnitude similar to the measurement uncertainty cannot be reliably established from the data. 
In Table~\ref{tab:negatives} we see that the predicted flux drop was smaller than, or equal to,
the measurement uncertainty for the events 200807, 220907 and  211007.
In contrast, the measurement uncertainties were 40\% and 20\% of the smallest predicted drop in the 260707 and 060807
events respectively. Hence, those last two events do contain information on the satellites' position and 
future ephemerides of the satellites must be able to reproduce these negative results.  

It is interesting that these two cases are partial events involving Miranda. Taking our positive observations of total eclipses
into account, a possible interpretation is that Miranda is out of position by an amount sufficient to cause a miss of a partial
event, but not of a total one. Such an effect may, for example, be caused by an offset in the inclination and longitude of the ascending node
of that satellite. Future processing of all the mutual event data amassed during the past Uranian equinox should determine if this 
is indeed the case.  

\section{Conclusions}
In this paper we have presented the results of a multi-instrument, multi-group effort
as part of a broader international campaign to capture the never-before-observed mutual events between the 
satellites of the planet Uranus. Eight such events, four eclipses and four occultations, were successfully detected
and reduced to precise intersatellite positions. Three of the eclipses reported involve the faint satellite Miranda, 
traditionally a difficult target for conventional astrometry due to its proximity to Uranus. Two additional observations yielded negative results above 
the noise, at variance with the predictions. The result of our data reduction procedure is a set of high-precision intersatellite 
positions. Our formal uncertainties are several times smaller than typical values for post-Voyager ground-based satellite-satellite 
astrometry \citep{Jones.et.al.1998,VeigaVieiraMartins1999,Shen.et.al.2002}. It is also noteworthy that, within the limitations of small number statistics, 
our rate of success was largely independent of the aperture used, 50\% in all cases. In our opinion, thorough planning 
(eg matching specific events to individual instrument capabilities) and instrument/detector operational flexibility were the main contributing factors
to the successful outcome of this programme. In particular, it highlights the ability of moderate aperture instruments to
carry out challenging observations.     

\begin{acknowledgements}
The Faulkes Telescope Project is an educational and research arm of LCOGT. FL acknowledges support from the 
Dill Faulkes Educational Trust.
Some of the observations reported in this paper were obtained with the Southern African Large Telescope (SALT), 
a consortium consisting of the National Research Foundation of South Africa, Nicholas Copernicus Astronomical 
Center of the Polish Academy of Sciences, Hobby Eberly Telescope Founding Institutions, Rutgers University, 
Georg-August-Universit\"{a}t G\"{o}ttingen, University of Wisconsin - Madison, Carnegie Mellon University, 
University of Canterbury, United Kingdom SALT Consortium, University of North Carolina - Chapel Hill, Dartmouth College, 
American Museum of Natural History and the Inter-University Centre for Astronomy and Astrophysics, India. 
Astronomical research at the Armagh Observatory is funded by the Northern Ireland Department of Culture, Arts and Leisure (DCAL). 
\end{acknowledgements}
\bibliographystyle{aa}
\bibliography{aa11523}

\begin{thebibliography}{19}
\expandafter\ifx\csname natexlab\endcsname\relax\def\natexlab#1{#1}\fi

\bibitem[{{Aksnes}(1974)}]{Aksnes74}
{Aksnes}, K. 1974, Icarus, 21, 100

\bibitem[{{Aksnes} \& {Franklin}(1976)}]{AksnesFranklin76}
{Aksnes}, K. \& {Franklin}, F.~A. 1976, AJ, 81, 464

\bibitem[{{Arlot} {et~al.}(2008){Arlot}, {Dumas}, \&
  {Sicardy}}]{Arlot.et.al.2008}
{Arlot}, J.-E., {Dumas}, C., \& {Sicardy}, B. 2008, A\&A, 492, 599

\bibitem[{{Arlot} {et~al.}(2006){Arlot}, {Lainey}, \&
  {Thuillot}}]{Arlot.et.al.2006}
{Arlot}, J.-E., {Lainey}, V., \& {Thuillot}, W. 2006, A\&A, 456, 1173

\bibitem[{{Birlan} {et~al.}(2008){Birlan}, {Nedelcu}, {Lainey}, {Arlot},
  {Binzel}, {Bus}, {Rayner}, {Thuillot}, {Vaduvesku}, \&
  {Colas}}]{Birlan.et.al.2008}
{Birlan}, M., {Nedelcu}, D.~A., {Lainey}, V., {et~al.} 2008, Astron.~Nachr.,
  329, 567

\bibitem[{{Christou}(2005)}]{Christou05}
{Christou}, A.~A. 2005, Icarus, 178, 171

\bibitem[{{Emelyanov} \& Gilbert(2006)}]{EmelyanovGilbert2006}
{Emelyanov}, N.~V. \& Gilbert, R. 2006, A\&A, 453, 1141

\bibitem[{{Giorgini} {et~al.}(1996){Giorgini}, {Yeomans}, {Chamberlin},
  {Chodas}, {Jacobson}, {Keesey}, {Lieske}, {Ostro}, {Standish}, \&
  {Wimberly}}]{Giorgini.et.al.1996}
{Giorgini}, J.~D., {Yeomans}, D.~K., {Chamberlin}, A.~B., {et~al.} 1996, BAAS,
  28, 1158

\bibitem[{{Hidas} {et~al.}(2008){Hidas}, {Christou}, \&
  {Brown}}]{Hidas.et.al.2008}
{Hidas}, M.~G., {Christou}, A.~A., \& {Brown}, T.~B. 2008, MNRAS, 384, L38

\bibitem[{{Jones} {et~al.}(1998){Jones}, {Taylor}, \&
  {Williams}}]{Jones.et.al.1998}
{Jones}, D.~H.~P., {Taylor}, D.~B., \& {Williams}, I.~P. 1998, A\&AS, 130, 77

\bibitem[{{Karkoschka}(2001)}]{Karkoschka01}
{Karkoschka}, E. 2001, Icarus, 151, 51

\bibitem[{{Lainey}(2008)}]{Lainey2008}
{Lainey}, V. 2008, P\&SS, 56, 1766

\bibitem[{{Laskar} \& {Jacobson}(1987)}]{LaskarJacobson87}
{Laskar}, J. \& {Jacobson}, R.~A. 1987, A\&A, 188, 212

\bibitem[{{Miller} \& {Chanover}(2008)}]{MillerChanover08}
{Miller}, C. \& {Chanover}, N.~J. 2008, 40th DPS Meeting, Cornell U., Ithaca,
  NY, 11-15 October 2008, Poster~\#46.02

\bibitem[{{Noyelles} {et~al.}(2003){Noyelles}, {Vienne}, \&
  {Descamps}}]{Noyelles.et.al.2003}
{Noyelles}, B., {Vienne}, A., \& {Descamps}, P. 2003, A\&A, 401, 1159

\bibitem[{{Rush} \& {Jacobson}(2007)}]{RushJacobson07}
{Rush}, B. \& {Jacobson}, R.~A. 2007, American Astronomical Society, DDA
  meeting \#38, \#13.05

\bibitem[{{Shen} {et~al.}(2002){Shen}, {Qiao}, {Harper}, {Hadjifotinou}, \&
  {Liu}}]{Shen.et.al.2002}
{Shen}, K.~X., {Qiao}, R.~C., {Harper}, D., {Hadjifotinou}, K.~G., \& {Liu}, R.
  2002, A\&A, 391, 775

\bibitem[{{Vasundhara} {et~al.}(2003){Vasundhara}, {Arlot}, {Lainey}, \&
  {Thuillot}}]{Vasundhara.et.al.2003}
{Vasundhara}, R., {Arlot}, J.-E., {Lainey}, V., \& {Thuillot}, W. 2003, A\&A,
  410, 337

\bibitem[{{Veiga} \& {Vieira Martins}(1999)}]{VeigaVieiraMartins1999}
{Veiga}, C.~H. \& {Vieira Martins}, R. 1999, A\&AS, 138, 247

\end{thebibliography}
\end{document}